\documentclass[12pt]{iopart}
\usepackage[utf8]{inputenc}
\expandafter\let\csname equation*\endcsname\relax
\expandafter\let\csname endequation*\endcsname\relax

\usepackage{amsmath}
\usepackage{amsfonts}
\usepackage{etoolbox}
\usepackage{ulem}
\usepackage{graphicx}
\patchcmd{\subequations}{\alph{equation}}{\textit{\alph{equation}}}{}{}

\usepackage{cleveref}

\crefname{equation}{}{}

\begin{document}

\title{A Review of Lagrangian Formalism in Biology: Recent Advances 
and Perspectives}

\author{D.T. Pham$^1$ and Z.E. Musielak$^2$}

\address{$^1$Department of Biology, University of Texas at Arlington, Arlington, TX 76019, USA \\}

\address{$^2$Department of Physics, University of Texas at Arlington, Arlington TX 76019, USA \\}

\begin{abstract}
The Lagrangian formalism has attracted the attention of mathematicians 
and physicists for more than 250 years and has played significant roles 
in establishing modern theoretical physics.  The history of the Lagrangian 
formalism in biology is much shorter, spanning only the last 50 years. 
In this paper, a broad review of the Lagrangian formalism in biology is 
presented in the context of both its historical and modern developments.
Detailed descriptions of different methods to derive Lagrangians for five
selected population dynamics models are given and the resulting 
Lagrangians are presented and discussed.  The procedure to use the 
obtained Lagrangians to gain new biological insights into the evolution 
of the populations without solving the equations of motion is described 
and applied to the models.  Finally, perspectives of the Lagrangian 
formalism in biology are discussed.

\end{abstract}



\section{Introduction}

The calculus of variations was originally developed by 
Leonhard Euler [1] and by Joseph-Louis Lagrange [2] in the 
18th Century as a tool to solve problems of finding maxima 
and minima that were not treatable by elementary calculus.  
In physical applications, the calculus of variations is typically 
formulated by introducing a scalar function called Lagrangian, 
which represents the difference between the kinetic and potential 
energies of a system.  Then, the action is defined as the integral 
over the Lagrangian from an initial to final time.   In general, the 
action integral has a different value for every path connecting the 
initial and final points.  The principle of least action [2] is used to 
determine the path that corresponds to the lowest energy.  

In the 19th Century, W. R. Hamilton [3] generalized the principle 
of least action to the Principle of Stationary Action (PSA), also 
known as the Hamilton Principle, and demonstrated that the 
PSA can be used to find a path for which the value of the action 
integral is stationary.  Originally, the PSA was developed for 
classical mechanics, and it was shown that its formulation was 
equivalent to the Newton second law of dynamics (e.g., [4-7]), 
which means that the principle does not yield any new view 
of classical mechanics.  However, the PSA allows formulating 
other theories of physics, especially, those involving classical 
and quantum fields  (e.g., [8-10]).  Thus, the PSA is one of the 
most powerful and elegant principles of modern physics. 

The Principle of Stationary Action forms the basis for the 
Lagrangian formalism, which is a procedure used to derive
an equation of motion for a given dynamical system, whose
Lagrangian is known.  The procedure is straightforward and 
requires substituting the Lagrangian into the Euler-Lagrange 
equation (e.g., [4-7]), which is the necessary condition 
that the action integral has its stationary value.  After the 
substitution is made and the required derivatives are 
calculated, the resulting equation of motion guarantees 
that its solutions describe the shortest trajectory along 
which the considered system evolves in time.     

There are three families of Lagrangians: standard, non-standard 
and null Lagrangians.  In standard Lagrangians (SLs), the kinetic 
and potential energy-like terms are easily recognized, and the 
role of these Lagrangians in classical mechanics has been 
well-established (e.g., [2,4-7]).  However, non-standard 
Lagrangians (NSLs), originally introduced by Arnold [5], 
who called them 'non-natural', are characterized by terms 
that can neither be identified as the kinetic nor potential 
energy-like terms.  Nevertheless, both SLs and NSLs give 
the same equations of motion when used in the Lagrangian 
formalism.  The family of null Lagrangians (NLs) is very 
different from the other two families.  The main difference 
is that NLs are defined as the total derivative of any scalar 
function, called here a gauge function, and that they 
identically satisfy the Euler-Lagrange equation [11].  

The Lagrangian formalism is commonly used in theoretical 
physics to derive all its fundamental equations [8-10].  There
have been attempts to establish theoretical biology by developing 
mathematical models of some biological systems (e.g., [12]). 
Such attempts include applications of the Lagrangian formalism 
to theoretical biology, or more specifically to population dynamics, 
which plays a special role in understanding competition for 
resources and predation in complex biological communities,
and for preserving biodiversity [13].  The main purpose of this 
review paper is to describe the current status of the Lagrangian 
formalism in theoretical biology and discuss its perspectives of 
becoming the main method to derive fundamental equations 
of theoretical biology.

The paper is organized as follows: The Lagrangian formalism for 
biologists and different families of Lagrangians are described in 
Section 2; population dynamics models considered in the paper
and their first-order and second-order equations of motion are 
given in Sections 3 and 4, respectively; standard and non-standard 
Lagrangians for the models are presented and discussed in Sections
5 and 6, respectively; Section 7 contains null Lagrangians and their 
gauge functions; applications of the obtained Lagrangians are 
considered and discussed in Section 8; perspectives of the 
Lagrangian formalism in biology are described in Section 9; 
and Conclusions are given in Section 10.

\section{Lagrangian formalism for biologists}

\subsection{Principle of stationary action}

The term action was originally introduced by Maupertuis in 1746
and then used to formulate a general principle of least action that 
would allow finding a local equation of motion for a given dynamical 
system [1-8].  One form of action was an integral over the distance 
along a given curve.  Euler [1] was the first to use it in 1744
and derived the condition for minimizing the integral.  However, 
later it became clear that instead of minimizing the integral, it 
was necessary to find its stationary value.  Thus, the Principle 
of Stationary Action (PSA) was formulated by Hamilton [3] in 
1834, and it is typically stated in the following form:

\bigskip
\noindent
{\it Of all the possible paths along which a dynamical system may 
move from one point to another in configuration space within a 
specified time interval, the actual path followed is that which 
minimizes the time integral of the Lagrangian function for the 
system.}\\
\noindent

To express the PSA mathematically, the action $\mathcal{A} 
(x, t_i, t_f)$ between the initial, $t_i$, and final, $t_f$, values of 
time, and with $x$ being a dependent variable representing a 
physical system, is defined as 
\begin{equation}
\mathcal{A} (x, t_i, t_f = \int_{t_i}^{t_f} L [\dot x (t), x (t), t] dt\ ,
\label{s2eq1}
\end{equation}
where $L [\dot x (t), x (t), t]$ is a Lagrangian and $\dot x$ is a 
derivative of $x$ with respect to $t$.  Then, the PSA requires 
that $\delta \mathcal{A} = 0$, where $\delta$ is the variation 
known also as the functional (or Fr\'echet) derivative of 
$\mathcal{A}$ with respect to $x(t)$.  Using $\delta \mathcal
{A} = 0$, the following Euler-Lagrange (E-L) equation is obtained
\begin{equation}
\left [ {d\over{dt}}\left({{\partial}\over{\partial{\dot x}}}\right) - 
{{\partial}\over{\partial x}} \right ] L [\dot x (t), x (t), t] = 0\ ,
\label{s2eq2}
\end{equation}
which can also be written as $\hat {EL} [ L (\dot x, x, t) ] = 0$,
where $\hat {EL}$ is the E-L operator defined by the derivatives 
in the square bracket in the above equation.  The E-L equation 
is the necessary condition for the action to be stationary, or to 
have either a minimum or maximum or saddle point.  The 
described procedure is the basis for the Lagrangian formalism
that is commonly used in modern physics to derive the equations 
of motion for a variety of physical systems (e.g., [8], and references
therein).  

The Lagrangian formalism is used to obtain an equation of motion
for a given dynamical system; however, this requires prior knowledge 
of a Lagrangian.  Once the Lagrangian is known, the process of 
finding the resulting dynamical equation is straightforward and 
it requires substitution of this Lagrangian into the E-L equation.  
For dynamical systems, whose total energy is conserved, the 
existence of Lagrangians is guaranteed by the Helmholtz conditions 
[14], which can also be used to obtain Lagrangians.  The procedure 
of finding Lagrangians for given equations of motion is called the 
inverse (or Helmholtz) problem of calculus of variations [15-17].
Different methods are required to solve the Helmholtz problem 
for the SLs, NSLs and NLs, and the main goal of this review paper
is to describe such methods and apply them to specific biological 
problems for which the equations of motion are already known. 

\subsection{Standard Lagrangians}

As originally demonstrated by Lagrange [2], for one-dimensional 
dynamical systems Lagrangians can be given in the following form
\begin{equation}
L_s [\dot x (t), x (t)] = E_{kin} (\dot x) - E_{pot} (x)\ ,
\label{s2eq3}
\end{equation}
where $E_{kin} (\dot x) = m \dot x^2 / 2$ is the kinetic energy, 
with $m$ being mass, and $E_{pot}$ is the potential energy, 
whose specific form depends on a considered dynamical system.  
Comparison of $L_s [\dot x (t), x (t)]$ to $L [\dot x (t), x (t), t]$
shows that the former does not depend explicitly on time.  The 
Lagrangians in which the kinetic and potential energy terms can
be identified are called {\it standard Lagrangians} (SLs); such 
Lagrangians are used to derive equation of motion for conservative 
dynamical systems (e.g., [18-21]).  

The Lagrangian formalism based on standard Lagrangians is 
commonly used in classical mechanics to obtain the law of inertia,
and the equations of motion for undriven and undamped harmonic 
oscillators, a linear and undamped pendulum, and other dynamical 
systems (e.g., [4-7]).  The Lagrangians required by the formalism  
are found by either guessing, or using the invariance of a physical 
system under consideration, or the structure of its already given 
equation of motion.  There are also other methods to derive SLs
for one-dimensional dynamical systems described by second-order 
ordinary differential equations (ODEs) (e.g., [18-22]).  A general 
method to obtain the SLs for population dynamical systems was 
developed in [23], and this method is described in detail in this 
review paper in Section 4.1.

\subsection{Non-standard Lagrangians}

The non-standard Lagrangians (NSLs) were originally introduced 
by Arnold [5], who called them 'not-natural' Lagrangians as he 
considered the SLs to be 'natural' Lagrangians.  The main 
difference between the SLs and NSLs is that in the latter the 
kinetic and potential energy-like terms are not easily distinguishable.  
In general, there are two classes of the NSLs, those that depend 
explicitly on time and those that are not explicit functions of time.  
The NSLs of these two classes can be written as 
\begin{equation}\
L_{ns1}[\dot{x}(t),x(t),t]={\frac{1}{f (t)\dot{x}(t)+g(t)x(t)+h(t)}}\ ,
\label{s2eq4}
\end{equation}
where $f(t)$, $g(t)$ and $h(t)$ are arbitrary and at least twice 
differentiable scalar functions of the independent variable $t$, 
and 
\begin{equation}\
L_{ns2}[\dot{x}(t),x(t)]={\frac{1}{F(x)\dot{x}(t)+G(x)x(t)+H(x)}}\ ,
\label{s2eq5}
\end{equation}
where $F(x)$, $G(x)$ and $H(x)$ are arbitrary and at least twice 
differentiable scalar functions of the dependent variable $x$.  Despite 
the fact that $L_{ns2}[\dot{x}(t),x(t)]$ does not depend explicitly 
on time, this Lagrangian is called non-standard because of its form, 
which neither resembles kinetic nor potential energy-like terms.  

The Lagrangians $L_{ns1}[\dot{x}(t),x(t),t]$ and $L_{ns2}[\dot{x}
(t),x(t)]$ are applicable to the following respective equations of motion
\begin{equation}
\ddot x+a(t)\dot x^2+b(t)\dot x+c(t)x= 0\ ,
\label{s2eq6}
\end{equation}
and
\begin{equation}
\ddot x+\alpha (x)\dot x^2+\beta(x)\dot x+\gamma(x)x=0\ .
\label{s2eq7}
\end{equation}
The coefficients $a(t)$, $b(t)$, $c(t)$, $\alpha(x)$, $\beta(x)$ and $\gamma(x)$
are at least twice differentiable functions, and they are given by the form of 
the equation of motion.  The functions $f(t)$, $g(t)$ and $h(t)$ are expressed 
in terms of the coefficients $a(t)$, $b(t)$ and $c(t)$ by substituting $L_{ns1} 
(\dot x,x,t)$ into the E-L equation [20,21].  However, the functions $F(x)$, 
$G(x)$ and $H(x)$ are determined by substituting $L_{ns2} (\dot x,x)$ into 
the E-L equation, which allows for these functions to be expressed in terms 
of the coefficients $\alpha(x)$, $\beta(x)$, and $\gamma(x)$ [20].   

Several different methods to derive the Lagrangian $L_{ns1}[\dot{x}(t),x(t),t]$ 
were developed and applied to different dynamical systems, whose equations 
of motion were already known (e.g., [20,21,24,25]).  However, studies of 
$L_{ns2}[\dot{x}(t),x(t)]$ show limitations of this NSL to ODEs with their 
coefficients being functions of the dependent variable $x$ (e.g.,[20,21,26-31]).  
Since the NSLs of the form of $L_{ns2} [\dot{x}(t),x(t)]$ play important roles 
in the population dynamics, they are presented following [32] in Section 4.2.

An interesting class of NSLs are the Caldirola-Kanai Lagrangians [33-35], 
which are essentially SLs multiplied by an exponential function of time.
However, based on the definition of SLs given in Section 2.2, the explicit 
dependence on time makes these Lagrangians non-standard; a method 
to derive the Caldirola-Kanai Lagrangians is presented in [36].  There is 
also an important class of NSLs that are called the El-Nabulsi Lagrangians 
[37], which are exponential, logarithmic, or different power-laws of SLs;
numerous applications of these Lagrangians to a diverse variety of physical 
problems have shown that they have more universal applications than SLs 
(e.g., [37-42]).

A powerful method to obtain NSLs of different forms was developed by 
using the Jacobi last multiplier [43], and the method was successfully 
applied to a variety of physical systems described by linear and 
nonlinear ODEs as well as to some mathematical (Riccati) equations 
(e.g., [44-54]).  The method was also used to derive NSLs for several 
population dynamics models [49,50], thus, the method and the obtained 
results are described in detail in this review paper (see Section 4.2).

A very different NSL was constructed for a simple harmonic oscillator 
in [55], which was later used to construct another NSL for a nonlinear 
oscillator [56].  The basic idea is that one NSL found for a linear oscillator
can be used to construct a new NSL for a nonlinear oscillator, and that 
the latter reduces to the former when the nonlinearity is neglected in 
the equation [56].  There is also a novel method to construct NSLs 
from null Lagrangians [57,58]; however, this method is presented 
after null Lagrangians are introduced. 

\subsection{Null Lagrangians} 

All Lagrangians that belong to the family of null Lagrangians (NLs) 
have two main characteristics: they identically yield zero when the E-L 
operator is acting on them  $\hat {EL} [ L_{null} (\dot x, x, t) ] = 0$,
and they always have corresponding gauge functions.  These gauge 
functions are scalar functions and their total derivatives are NLs.  As
a result, NLs do not contribute to the equations of motion that are 
derived from the E-L equation (see Eq. \ref{s2eq2}), which may 
imply that NLs do not play any important role either in physics or
biology.  However, NLs have been studies in mathematics (e.g., 
[11,59-65]), and there are also their several selected applications 
in physics (e.g., [66-71]); in biology, first NLs were determined 
for some population dynamics models in [32] and their description 
is given in Section 4.3.

A method that uses NLs to obtain NSLs was developed in [57],
and it was demonstrated that substitution of any NL into $d 
L_{null} / dt = 0$, which can be explicitly written as 
\begin{equation}\
         \frac{d L_{null}(\dot x, x,t)}{dt} = 
     \frac{\partial L_{null}(\dot x, x,t)}{\partial t} + \dot x 
     \frac{\partial L_{null}(\dot x, x,t)}{\partial x} + \ddot x
     \frac{\partial L_{null}(\dot x, x,t)}{\partial \dot x} = 0\ ,
\label{s2eq8}
\end{equation}
gives an equation of motion, where $L_{null} = d \Phi / dt$, 
with $\Phi (x,t)$ being a gauge function.  This shows that 
Eq. (\ref{s2eq8}) plays the same role for NLs as the E-L 
equation plays for SLs and NSLs.  However, the resulting 
equation of motion is limited because its coefficients are 
required to obey special relationships, which are different 
for different equations.

Recent studies of NLs [58] shown that the previous work [57] 
can be generalized and that for some dynamical systems, this 
generalization allows deriving equations of motion without any 
additional limitations.  Moreover, the studies also demonstrated 
that the inverse of any NL generates a NSL, whose substitution 
into the E-L equation gives a new equation of motion.  This is 
an interesting result that shows the close relationship between 
NLs and NSLs.  The relationship allows for the construction of 
new NSLs for all known NLs, but also for finding NLs for known 
NSLs; the latter was used in [32] to obtain first NLs for the 
population dynamics models. 

\subsection{First attempts to establish Lagrangian formalism in biology} 

The original idea of using a Lagrangian in theoretical biology was 
proposed by Kerner [72], who utilized the Legendre transformations 
to find a Hamiltonian and demonstrated that the obtained Hamiltonian 
was an integral of motion.  A general canonical procedure of constructing 
Lagrangians for systems of coupled first-order ODEs was developed by 
Trubatch and Franco [73], who based their method on the earlier work 
of Helmholtz [14] and Havas [74].  A different approach was considered 
by Paine [75], who used the Helmholtz conditions [14] to investigate the 
existence and construction of Lagrangians for the set of ODEs examined 
by Kerner, and to demonstrate importance of variational self-adjointness 
in the process. 

Among a large variety of biological systems, the authors selected 
several population dynamics models, whose equations of motion 
allowed finding their Lagrangians [73,75].  The specific considered 
models were the Lotka-Volterra [76-80], Verhulst [81,82], Gompertz 
[83,84] and Host-Parasite [85] models.  The obtained Lagrangians 
[73,75] must be considered as the NSLs because neither kinetic nor 
potential energy-like terms can be identified, and they also show 
explicit dependence on time.  Nevertheless, the found Lagrangians 
allow deriving the equations of motion for the population dynamics 
models after they are substituted into the E-L equation. 

The choice of the four population dynamics models among a large 
variety of biological systems was justified by the fact that the population 
dynamics is the key to understanding the competition for resources 
and predation in complex communities, which becomes essential in 
preserving biodiversity (e.g., [13,86-88]).  In recent work described 
in the remaining sections of this paper, the Lagrangian formalism in 
population dynamics is established by developing methods to construct 
the SLs, NSLs and NLs for the considered models.  However, it must 
be pointed out that the presented methods can also be used to find 
Lagrangians for a broad range of biological systems whose equations
of motion are already known.   

\section{Population dynamics models and first-order 
equations of motion}

\subsection{Symmetric models} 

Models of population dynamics can be classified as symmetric and 
asymmetric, where being symmetric means that the dependent 
variables and the constant coefficients in the equations of motion 
for these models can be replaced by each other; on the other 
hand, such replacements cannot be done for asymmetric models.  
In selecting the models, we follow [13,22,32,48,49,74] and 
consider the following three symmetric models: the Lotka-Volterra 
[76-80], Verhulst [81,82] and Gompertz [83,84] models that 
describe two interacting (preys and predators) species.  The 
interaction is represented mathematically by two coupled,
damped and nonlinear first-order ordinary differential 
equations (ODEs) as shown in Table 1 (e.g., [86-88]). 

\renewcommand{\arraystretch}{1.1}
\begin{center}\begin{tabular}
{ll} \hline
{\bf Population models}	&{\bf Equations of Motion}\\ \hline
Lotka-Volterra Model &	$\dot{w_1}=w_1(a\:+\:bw_2\:)$\\
			&	$\dot{w_ 2}=w_2(A\:+Bw_1)$\\ \hline
Verhulst Model 	&	$\dot{w}_1=w_1(A+\:Bw_1\:+f_1 w_2)$\\
			&	$\dot{w}_2=w_2(\:a+\:bw_2\:+f_2 w_1)$\\ \hline
Gompertz Model 	&	$\dot{w}_1=w_1{(}A\log{\bigr(}{w_1\over m_1}
{\bigr)}+B{w_2}{)}$\\ 
			&	$\dot{w}_2=w_2{(}\:a\log{\bigr(}{w_2\over m_2}
{\bigr)}+b{w_1})$\\ \hline   
\end{tabular}\end{center}

Table 1. Equations of motion for the symmetric population dynamics models\\

The dependent variables $w_1 (t)$ and $w_2 (t)$ in these models 
describe prey and predators of the interacting species, respectively. 
However, the time derivatives $\dot w_1 (t)$ and $\dot w_2 (t)$ show 
how the population of these species varies in time.  The coefficients 
$a$, $b$, $A$, $B$ $f_1$, $f_2$, $m_1$ and $m_2$ are constant, 
and they must be real because they represent the interaction between 
the species; their detailed description is given in Section 4.2.

\subsection{Asymmetric models} 

Among the asymmetric population dynamics models, we consider two 
different models, namely, the Host-Parasite [86] that describes two 
interacting (preys and predators) species, and the SIR model [89,90]
that is used to represent the spread of a disease in a given population. 
Similarly, to the symmetric models, the variables $w_1(t)$ and $w_2(t)$
in the Host-Parasite model describe prey and predators of the interacting 
species, respectively; note differences in the mathematical forms of the 
symmetric and asymmetric equations.  On the other hand, in the SIR 
model, the variables $w_1(t)$ and $w_2(t)$ describe susceptible and 
infectious populations.  The differences in the mathematical 
structure of the ODEs representing these two asymmetric models 
must be noted.  Moreover, the differences will be also be reflected
in the Lagrangians derived for these two asymmetric models.  
\renewcommand{\arraystretch}{1.1}
\begin{center}\begin{tabular}
{ll} \hline
{\bf Population models}	&{\bf Equations of Motion}\\ \hline
Host-Parasite Model	&	$\dot{w}_1=w_1(a\:-bw_2)$\\ 
			&	$\dot{w}_2=w_2(A-B{w_2\over w_1})$ \\ \hline
SIR Model 		&	$\dot w_1=-b w_1w_2$\\
			&	$\dot w_2=\:\:\:b w_1w_2-aw_2$\\ \hline   
\end{tabular}\end{center}

Table 2. Equations of motion for the asymmetric population dynamics models\\

The coefficients $a$, $b$, $A$ and $B$ are real constants and they describe
the interaction between the species; their detailed description can be found 
in Section 4.3.

The first-order ODEs given in Tables 1 and 2 can be solved to determine 
the changes in the populations of both species resulting from their mutual
interactions (e.g., [86,88]).  However, the main emphasis of this review 
paper is not on the time-dependent solutions to the ODEs, but instead 
on establishing the Lagrangian formalism that allows finding Lagrangians
that give equations of motion when substituted into the E-L equation.

\section{Population dynamics models and second-order 
equations of motion}

\subsection{General equation of motion}

The population dynamics models given in Tables 1 and 2 can also 
be expressed mathematically as coupled, damped and nonlinear 
second-order ODEs.  The choice between the first-order or second-order
ODEs being used to describe the population dynamics is motivated 
by the information available; e.g., if only one population is observable
but the other one is unknown.   Thus, the first-order ODEs from Table
1 and 2 can be cast into the following general second-order equation 
of motion
\begin{equation}
\ddot w_{i}+\alpha_{i}(w_{i})\dot w_{i}^2+\beta_{i}(w_{i})
\dot w_{i} + \gamma_{i}(w_{i})w_{i} = F_{i}(\dot w_i, w_i)\ ,
\label{s2eq8}
\end{equation}
where $i=1$ and $2$, which means that for each population 
dynamics model, two separate equations are obtained, one 
for $w_1$ and another for $w_2$.  This equation generalizes 
Eq. (\ref{s2eq7}) by allowing for a driving force to be included,
thus, it is the general equation of motion for the considered 
population dynamics models.  It must be noted that this 
general equation of motion does not depend explicitly on 
time, despite the fact 

The general equation of motion contains the terms that can 
be identified as damping and nonlinearity, and also the driving 
force term, and the non-constant coefficients $\alpha_{i}(w_{i})$, 
$\beta_{i}(w_{i})$ and $\gamma_{i}(w_{i})$, and the driving 
force $F_{i}(\dot w_i, w_i)$ are different for different population 
dynamics models; their explicit forms are given in Section 5.

The presence of $F_{i}(\dot w_i, w_i)$ in the general equation 
of motion requires that the E-L equation is also modified and 
written in the following form
\begin{equation}
{d\over dt}\biggr({\partial L\over\partial\dot w_i}\biggr)-{\partial L
\over\partial w_i} =  F_{i}(\dot w_i, w_i)\ .
\label{s2eq9}
\end{equation}
Any Lagrangian obtained for Eq. (\ref{s2eq8}) must be substituted
into Eq. (\ref{s2eq9}), so the correct forcing term appears in the 
resulting equation of motion.  This is a well-known procedure that 
can be found in standard textbooks of classical mechanics (e.g., 
[6,7]). 

In a special case when $\alpha_{i} = 0$, $\beta_{i} = b_i$ 
= const, $\gamma_{i} = c_i$ = const, and $F_{i}$ is an 
explicit and periodic function of $t$, Eq. (\ref{s2eq9}) may 
show chaos for certain values of these coefficients (e.g., 
[90,91]).  However, no chaos has yet been observed in the 
population dynamics models (e.g., [13,86-88]).  The main 
reason could be the lack of the force periodic in time force 
in these systems since the form of $F_{i}(w_i)$ in the 
population dynamics models is very different from that 
known for chaotic systems in classical dynamics (e.g., 
[91,92]).  It must be pointed out that some studies 
suggested that insect population can undergo transitions 
between stable and chaotic phases for some population 
dynamics models (e.g., [93,94]).

\subsection{Equations of motion for symmetric models} 

The first-order equations of motion for the symmetric models 
given in Table 1 are now cast into the second-order equations 
of motion for the dependent variables $w_1 (t)$ and $w_2 (t)$; 
the physical meaning of the nonconstant coefficients in the 
resulting equations of motion is also explained.

For the Lotka-Volterra model, the resulting equations of motion 
are given by 
 \begin{equation}
\ddot w_1 - {1\over w_1} \dot w_1^2 - (Bw_1+A)\dot w_1 + 
(Bw_1+A) a w_1 = 0\ ,
\label{s2eq10a}
\end{equation}  
and
\begin{equation}
\ddot w_2 - {1\over w_2} \dot w_2^2 - (\:bw_2+a)\dot w_2 +
 (b w_2+a) A w_2 = 0\ ,
\label{s2eq10b}
\end{equation}
which explicitly show the presence of linear and nonlinear damping 
terms and no driving force.  The coefficients in these equations represent: 
$a$ is reproduction rate of prey, $b$ is mortality rate of predator per prey, 
$A$ is predation rate of predator, and $B$ is reproduction rate of predator 
per prey.  The obtained equations of motion have both the linear and 
nonlinear damping terms but show no forcing functions.

The Verhulst model describes the growth of a single population
with self-interaction damping, which accounts for the carrying
capacity of the environment.  The coefficients in these equations 
represent: $a$ is growth rate, $b$ is the self-interaction term, $A$ 
is mortality, and $B$ is carrying capacity.  Thus, the second-order 
equations of motion for the dynamical variables of the Verhulst 
model are:

\[
\ddot w_1 - (1 + b) {1\over w_1} \dot w_1^2 + \bigr[(2b-1)Bw_1
- f_2w_1^2+(2Ab-a)+(f_2-b)B\bigr] \dot w_1
\]
\begin{equation}
+ [ \bigr(f_2-b\:\bigr)Bw_1^2
+\bigr(Af_2-2Ab-a)w_1+A\bigr(a-Ab\bigr)] w_1 = 0\ ,
\label{s2eq11a}
\end{equation}  
and
\[
\ddot w_2 - (1 + B) {1\over w_2} \dot w_2^2 + 
\bigr[(2B-1)bw_2-f_1w_2^2+(2aB-A)+(f_1-B)b\bigr] \dot w_2
\]
\begin{equation}
+ [ \bigr(f_1-B\bigr)bw_2^2+
\bigr(\:af_1-2aB-A)w_2+a\bigr(A-aB\bigr) ] w_2 = 0\ ,
\label{s2eq11b}
\end{equation}
where the constant $f_1$ represents the population intrinsic growth 
rate at which the population would grow if there were no constraints 
on resources, and the constant $f_2$ is the carrying capacity of the 
environment that corresponds to the maximum population size that 
the environment can sustain indefinitely. 

The Gompertz model is another population model describing the 
growth dynamics by an alternative self-interaction damping term.
Therefore, the physical meanings of the coefficients $a$, $b$, $A$ 
and$B$ are the same as in the Verhulst model.  The difference 
between the two models is that the coefficient $B$ in the 
Gompertz indirectly limits the population size, describing a 
more gradual, asymmetric approach to the carrying capacity. 
However, the resulting second-order equations of motion for 
the dynamical variables of the Gompertz model are:
 \begin{equation}
\ddot w_1 - {1\over w_1} \dot w_1^2 - (Am_1+\:bw_1) 
\dot w_1 + \bigr[A\log\bigr({w_1\over m_1}\bigr)\bigr] w_1^2 
= F_1(\dot w_1,w_1)\ ,
\label{s2eq12a}
\end{equation}  
and
\begin{equation}
\ddot w_2 - {1\over w_2} \dot w_2^2 - (am_2+\:Bw_2) 
\dot w_2 + \bigr[a\:\log\bigr({w_2\over m_2}\bigr)\bigr]
w^2_2 = F_2(\dot w_2,w_2)\ ,
\label{s2eq12b}
\end{equation}
where the forcing terms are given by $F_1(\dot w_1,w_1)=
(\dot w_1 - A w_1) g_1(\dot w_1,w_1)$ and $F_2(\dot w_2,w_2)
=(\dot w_2 - a w_2) g_2(\dot w_2,w_2)$, with 
\begin{equation}
g_1(\dot w_1,w_1)\:=\:a\log\:\biggr[{1\over m_2B}\:\biggr({\dot w_1\over w_1}-
A\log\biggr({w_1\over m_1}\biggr)\biggr)\biggr],
\label{s2eq13a}
\end{equation}
and 
\begin{equation}
g_2(\dot w_2,w_2)\:=A\log\:\biggr[{1\over m_1b}\:\:\biggr({\dot w_2\over w_2}-
a\log\biggr({w_2\over m_2}\biggr)\biggr)\biggr].
\label{s2eq13b}
\end{equation}
The coefficient $m_1$ is the scaling constant that sets the initial offset from 
the carrying capacity, and $m_2$ is the growth rate constant that determines 
the rate at which the population approaches the carrying capacity.  

The obtained equations of motion have both the linear and nonlinear damping 
terms as well as the driving force terms that are functions of the dependent 
variables and their time derivatives.  The driving forces $F_1(\dot w_1,w_1)$ 
and $F_2(\dot w_2,w_2)$ contain the functions $g_1(\dot w_1,w_1)$ and 
$g_2 (\dot w_2,w_2)$ from which the variables $\dot w_1$ and $\dot w_2$ 
cannot be separated, and thus they cannot be included in the linear and 
nonlinear damping terms on the LHS of Eqs (\ref{s2eq13a}) and 
(\ref{s2eq13b}).   

By comparing the above equations of motion to Eq. (\ref{s2eq8}),
the explicit forms of the nonconstant coefficients $\alpha_{i}(w_{i})$,
$\alpha_{i}(w_{i})$ and $\alpha_{i}(w_{i})$ can be found for the  
three symmetric population dynamics models given in Table 1.

\subsection{Equations of motion for asymmetric models} 

The first-order equations of motion for the asymmetric models 
given in Table 2 are now cast into the second-order equations 
of motion for the dependent variables $w_1 (t)$ and $w_2 (t)$, 
and the nonconstant coefficients in these equations are defined.

The Host-Parasite model describes the interaction between a host 
and its parasite by taking into account the nonlinear effects of the 
host population size on the growth rate of the parasite population. 
The second-order equations of motion are given by 
 \begin{equation}
\ddot w_1 - {1\over w_1}\Bigr(1+{B\over bw_1}\Bigr) \dot w_1^2 
+ (A-{2aB\over bw_1})\dot w_1 + a A w_1 = F_1\ ,
\label{s2eq14a}
\end{equation}  
and
\begin{equation}
\ddot w_2 - {2\over w_2} \dot w_2^2 + (bw_2-a-A)\dot w_2 + 
A(bw_2-a) w_2 = 0\ ,
\label{s2eq14b}
\end{equation}
where the coefficient $a$ represents the growth rate of the host, $b$ 
represents the interaction rate between the host and parasite, $A$ is 
the natural death rate of parasite, and $B$ describes the effect of 
parasite on the host.  In addition, the forcing term is $F_1 = B{\:\:
a^2\over b}$ = const.  The obtained equations of motion have both 
the linear and nonlinear damping terms; however, an interesting 
result is that the constant driving force is only present in the equation 
of motion for $w_1$.

The SIR model describes the spread of a disease in a population, and 
it is achieved by dividing the population into a susceptible population 
denoted by $w_1(t)$, the infectious population represented by $w_2(t)$, 
and a recovered population $w_3(t)$.  It is seen in Table 2 that the 
dependent variable $w_3(t)$ does not appear explicitly in the set of 
first-order ODEs describing mathematically this model.  The reason is 
that the population conservation law $d/dt(w_1 + w_2 + w_3)=0$ 
implies that the sum $w_1 + w_2 + w_3$ = const, which allows 
expressing $w_1$ as a known function of $w_2$ and $w_3$; in
other words, the variable $w_3$ can be eliminated. 

Then, the time evolution equations for the variables $w_1(t)$ and 
$w_2(t)$ for the SIR model can be written as
\begin{equation}
\ddot w_1 - {1\over w_1} \dot w_1^2 - (bw_1 - a)\dot w_1= 0\ ,
\label{s2eq15a}
\end{equation}  
and
\begin{equation}
\ddot w_2 - {1\over w_2} \dot w_2^2 - bw_2\dot w_2 + 
a b w^2_2 = 0\ ,
\label{s2eq15b}
\end{equation}
where the constant coefficients $a$ and $b$ represent the infection and 
recovery rates, respectively.  Moreover, if $a>0$ and $b>0$, then the 
terms with these coefficients in the equations of motion describe 
newly infected and recovered individuals, respectively.

Comparison of the above equations to the general equation of motion
given by Eq. (\ref{s2eq8}) allows finding the explicit expressions for 
the nonconstant coefficients $\alpha_{i}(w_{i})$, $\beta_{i}(w_{i})$ 
and $\gamma_{i}(w_{i})$ for the two symmetric population dynamics 
models given in Table 2.

\section{Standard Lagrangians for population dynamics models}

\subsection{Equation of motion}

The general equation of motion given by Eq. (\ref{s2eq8}) contains
the linear damping term $\beta_{i}(w_{i}) \dot w_{i}$ that identically 
satisfies the E-L equation (see Eq. \ref{s2eq2}), which makes it a null 
Lagrangian (see Section 2.4).  The presence of this term in the equations
of motion derived in Sections 4.2 and 4.3 imposes some limitations on 
the construction of SLs for the population dynamics models (e.g., [14-16,
20,21,33-36]).  A possible solution is to include the term into the 
driving force and write Eq. (\ref{s2eq8}) as
\begin{equation}
\ddot w_{i}+\alpha_{i}(w_{i})\dot w_{i}^2+ \gamma_{i}(w_{i})
w_{i} = {\cal F}_{i}(w_i,\dot w_{i})\ ,
\label{s3eq1}
\end{equation}
where ${\cal F}_{i}(\dot w_{i},w_i) = F_{i}(\dot w_i,w_i) - \beta_{i}
(w_{i})\dot w_{i}$, and the explicit forms of $\alpha_{i}(w_{i})$, 
$\beta_{i}(w_{i})$ and $\gamma_{i}(w_{i})$ from each model are 
found by comparing the equations of motion given in Section 4 to 
Eq. (\ref{s3eq1}).  The force ${\cal F}_{i}(\dot w_{i},w_i)$ is 
known as a dissipative force [6,7] because of its explicit dependence 
on $\dot w_{i}$, and its presence in the equation of motion implies 
that in the E-L equation given by Eq. (\ref{s2eq9}), the forcing term 
is replaced by ${\cal F}_{i}(\dot w_{i},w_i)$. 

\subsection{Construction of standard Lagrangians}

The original method to construct a standard Lagrangian for Eq. (\ref{s3eq1})
was developed in [22] and then generalized in [23].  The method requires
performing an integral transform that introduces a scalar function, whose 
existence removes the nonlinear damping term from Eq. (\ref{s3eq1}).
Then, the process of finding the SL is straightforward and gives the 
following Lagrangian
\begin{equation}
L_{s,i} (\dot w_i, w_i)=\left [ E_{kin,i} (\dot w_i) - E_{pot,i} (w_i) \right ] 
e^{2I_{\alpha,i}(w_i)}\ ,
\label{s3eq2}
\end{equation}
where $E_{kin,i} (\dot w_i) = \dot w_i^2 / 2$ and 
\begin{equation}
E_{pot,i} (w_i) = e^{-2I_{\alpha,i}(w_i)} \int \displaylimits_{}^{w_i} \tilde w_i 
\gamma_i (\tilde w_i) e^{2I_{\alpha,i} (\tilde w_i)} d\tilde w_i\ ,
\label{s3eq3}
\end{equation}
with 
\begin{equation}
I_{\alpha,i}(w_i)=\int\displaylimits_{}^{w_i} \alpha_i (\tilde w_i)d\tilde w_i\ .
\label{s3eq4}
\end{equation}
The derived $L_{s,i}(\dot w_i, w_i)$ is called standard because the kinetic 
and potential energy-like terms can be recognized and they do not depend 
explicitly on time $t$.  Physical meaning of $L_{s,i}(\dot w_i, w_i)$ becomes 
more clearer when the coefficients $\alpha_i$ and $\gamma_i$ are constants, 
which is the case for many physical oscillatory systems.  However, it must 
be noted that in the population dynamics models considered here, those 
coefficients are functions of $w_i$ that is a dependent variable being a 
function of time $t$.  Then, the SLs for the models must be obtained 
using the above formulas.

To obtain the equation of motion given by Eq. (\ref{s3eq1}),
the derived SL must be substituted into the E-L equation $\hat 
{EL}[ L_{s,i} (\dot w_i, w_i) ] = {\cal F}_{i}(w_i,\dot w_{i}) 
e^{2I_{\alpha,i}(w_i)}$, whose explicit form is  
\begin{equation}
{d\over dt}\biggr({\partial L_{s,i}\over\partial\dot w_i}\biggr)-{\partial L_{s,i}
\over\partial w_i} = {\mathcal F}_{i}(w_i,\dot w_{i}) e^{2I_{\alpha,i}(w_i)}\ .
\label{s2eq20}
\end{equation}
The presence of the term ${\mathcal F}_{i}(w_i,\dot w_{i}) e^{2I_{\alpha,i}
(w_i)}$ in the E-L equation is justified by the fact that this term does not 
arise from any potential [23].

\subsection{Derived standard Lagrangians}

The above procedure can be used to derive the SLs for the symmetric and 
asymmetric population dynamics models, and the obtained Lagrangians 
are given in Table 3.

\renewcommand{\arraystretch}{1.1}
\begin{center}
\begin{tabular}{@{}lll@{}} 
\hline
{\bf Models}	&{\bf Standard Lagrangians }\\ 
\hline
Lotka-Volterra  &	$L_{s,1} = (1/2) (\dot w_1 / w_1)^2 - a(Bw_1+A \ln|w_1|)$\\
			&	$L_{s,2} = (1/2)(\dot w_2 / w_2)^2 - A(bw_2+a \ln|w_2|)$\\ \hline
            
Verhulst$^*$  	    &	$L_{s,1} =  (1/2)[(\dot w_1 / w_1)^2 - C_{11} w_1^2 - 
                                                   C_{12} w_1 + C_{13}] w_1^{-2b}$\\ 
			&	$L_{s,2} =  (1/2)[(\dot w_2 / w_2)^2 - C_{21} w_2^2 -  
                                                   C_{22} w_2 + C_{23}] w_2^{-2B}$\\ \hline
Gompertz  	& 	$L_{s,1} =  (1/2) (\dot w_1 / w_1)^2 - A (\ln|w_1/m_1| - 1) w_1$\\ 
			&	$L_{s,2} =  (1/2)(\dot w_2 / w_2)^2 - a (\ln|w_2/m_2| - 1) w_2$\\ \hline 
   
Host-Parasite$^{**}$	&	$L_{s,1} = (1/2) ({\dot w_1 / w_1})^2 Ex(2B/b w_1) + aA Ei(2B/b w_1)$\\
			&	$L_{s,2} = (1/2) ({\dot w_1 / w_1})^2 w_2^{-2} - A [(a/2w_2) - b] w_2$\\ \hline 
   
SIR  		    &	$L_{s,1} = (1/2) ({\dot w_1 / w_1})^2$\\ 
	       	&	$L_{s,2} = (1/2) ({\dot w_1 / w_1})^2 - ab w_2$\\ \hline
\end{tabular}\end{center}%

$^*$ $C_{11} = (f_2 - b) B /(1 - b)$, $C_{12} = 2 (Af_2 - 2Ab - a) /(1 - 2b)$, $C_{13} =
A(a - Ab) / b$, 

$C_{21} = (f_1 - B) b /(1 - B)$, $C_{22} = 2 (af_1 - 2aB - A) /(1 - 2B)$ and
$C_{23} = a(A - aB) / B$

\smallskip

$^{**}$ $Ex(2B/b w_1) = e^{2B/b w_1}$ and $Ei(2B/b w_1) = \int^{2B/b w_1}_{\infty} 
(e^{\eta}/{\eta}) d \eta$.\\

Table 3. Standard Lagrangians for the population dynamics models 
(after Pham \& Musielak [23]).\\

According to the results of Table 3, the derived SLs have one common factor 
$(1/2) (\dot w_i / w_i)^2$, which is only modified for the Verhulst and 
Host-Parasite models.  The factors can be considered as 'kinetic energy-like'
terms and they reproduce both terms $\ddot w_i$ and $\frac{1}{w_i} \dot 
w_i^2$ in the equations of motion, when the SLs are substituted into the 
E-L equation.  The remaining parts in the SLs can be considered as 'potential
energy-like' terms and they are very different for different models.  The 
difference is especially prominent for the Host-Parasite model, whose SL
obtained for $w_1$ shows explicit dependence on the exponential integral
$Ei(2B/b w_1)$; the presence of such an integral in the Lagrangian allows 
defining a new class of SLs within the family of standard Lagrangians.   

The derived Lagrangians are defined as standard Lagrangians because
they have terms that can be identified as kinetic and potential energy-like
terms and also because they do not depend explicitly on time; the latter
is an important characteristic of Lagrangians defined in this paper as 
SLs.  The SLs presented in Table 3 are the most basic Lagrangians for 
the models as there are also alternative Lagrangians, which may have 
much more complicated forms and yet they give the same equations 
of motion; such Lagrangians are presented in the next section.   

\section{Non-standard Lagrangians for population dynamics models}

\subsection{Jacobi last multiplier method and its non-standard Lagrangians} 

In the original work by Trubatch and Franco [72], it was demonstrated that 
a system of two first-order ODEs given by 
\begin{equation}
\dot w_1 = \phi_1 (w_1, w_2, t)\ ,
\label{s4eq1a}
\end{equation}  
and
\begin{equation}
\dot w_2 = \phi_2 (w_1, w_2, t)\ ,
\label{s4eq1b}
\end{equation}
always admits the following general linear Lagrangian
\begin{equation}
L (\dot w_1, \dot w_2, w_1, w_2, t) = U_1 (w_1, w_2, t) \dot w_1 +
U_2 (w_1, w_2, t) \dot w_2 - V (w_1, w_2, t)\ ,
\label{s4eq2}
\end{equation}  
which must be classified as a non-standard Lagrangian because of its 
explicit dependence on time and no presence of the kinetic energy-like 
term in this Lagrangian.  The main advantage of this method is that 

As pointed out by Nucci and Tamizhmani [49], the method developed 
in [73] resembles the Jacobi last multiplier method, which has been 
extensively used to obtain Lagrangians for different dynamical systems 
(e.g., [44-53]).  The main advantage of the method developed in [73] 
is that it applies directly to the first-order ODEs describing the models
and that allows finding the NSLs for the symmetric population models;
however, its main disadvantage is that it fails to obtain NSLs for 
asymmetric models, such as the Host-Parasite model.  

Following [48], for a general second-order ODE given as $\ddot w_i = 
\Phi_i (\dot w_i, w_i, t)$, the Jacobi last multiplier is directly related 
the Lagrangian $L_i (\dot w_i, w_i, t)$ by 
\begin{equation}
M_i (\dot w_i, w_i, t) = \frac{\partial^2 L_i (\dot w_i, w_i, t)}
{\partial \dot w_i^2}\ ,
\label{s4eq3}
\end{equation}  
where $M_i (\dot w_i, w_i, t)$ is the Jacobi last multiplier that satisfies
\begin{equation}
\frac{d}{dt} (log M_i (\dot w_i, w_i, t)) + \frac{\partial \Phi_i (\dot w_i, w_i, t)}
{\partial \dot w_i} = 0\ .
\label{s4eq4}
\end{equation}  
Once $M_i (\dot w_i, w_i, t)$ is known, then the Lagrangian is obtained from 
\begin{equation}
L_i (\dot w_i, w_i, t) = \int \left ( \int M_i (\dot w_i, w_i, t) d \dot w_i \right ) 
+ f_1 (w_i, t) \dot w_i + f_2 (w_i, t)\ ,
\label{s4eq5}
\end{equation}  
with 
\begin{equation}
f_1 (w_i, t) = \frac{\partial G (w_i, t)}{\partial x}\ ,
\hskip0.15in {\rm and} \hskip0.15in f_2 (w_i, t) = 
\frac{\partial G (w_i, t)}{\partial t} + f_3 (w_i, t)\ ,
\label{s4eq6}
\end{equation}  
where $f_3 (w_i, t)$ must satisfies $\ddot w_i = 
\Phi_i (\dot w_i, w_i, t)$, and $G (w_i, t)$ is an 
arbitrary function.

The described Jacobi multiplier method was used by Nucci and Tamizhmani 
[49] to obtain the NSLs for the population dynamics models considered by 
Trubatch and Franco [73].  As expected, the same NSLs were derived 
including the NSL for the Host-Parasite model that failed in [73].  The 
obtained NSLs are presented in Table 3.    

\renewcommand{\arraystretch}{1.4}
\begin{center}\begin{tabular}
{ll} \hline
{\bf Models}	&{\bf Non-standard Lagrangians}\\ \hline

Lotka-Volterra$^*$ &	$L_{ns} = \log(w_1) \frac{\dot{w}_2}{w_2} - \log(w_2)
\frac{\dot{w}_1}{w_1} - 2(A \log(w_1) - a \log(w_2))$\\
& $\hskip0.3in - 2( Bw_1 - bw_2) + \frac{d}{dt} G(t, w_1, w_2)$\\ \hline

Verhulst$^*$ 	&  $L_{ns} = \exp(b_3 t) w_2^{b_2} w_1^{b_1 + 1} 
\dot{w_2}^{b_1 + 1} - w_2^{b_1} \dot{w_1}^{b_2 + 1}$\\
& $ \hskip0.3in - w_2^{b_2 + 1} w_1^{b_1 + 1} \left( 2 f_2 
w_1^{b_1 + 2} + 2 b w_2^{b_1 + 1} + 2 a (b_2 + 1) \right)$\\
& $ \hskip0.3in - w_2^{b_2 + 1} w_1^{b_1 + 1} \left ( b_3 
(b_1 + 1)(b_2 + 1) \right) + \frac{d}{dt} G(t, w_1, w_2)$ \\ \hline
   
Gompertz$^*$  &	$L_{ns} = \exp[{-(a + A)t}] \left[ \log(w_1) \left( 
\frac{\dot{w}_2}{w_2} \right) - \log(w_2) \left( \frac{\dot{w}_1}
{w_1} \right) \right]$\\ 
& $ \hskip0.3in - 2 \exp[{-(a + A)t}] \left [ a \log \left( \frac{w_2}{m_1} 
\right) \log(w_1) - A \log \left( \frac{w_1}{m_2} \right) \log(w_2)  \right ]$ \\
& $ \hskip0.3in + \exp[{-(a + A)t}] \left [ 2(Bw_2 + bw_1) - (A - a) \log(w_1) 
\log(w_2) \right]$ \\
& $ \hskip0.3in + \frac{d}{dt} G(t, w_1, w_2)]$ \\ \hline

Host-Parasite$^*$ &  $L_{ns} = \exp(At) \left [\log(w_1) \frac{\dot{w_2}}{w_2^2} + 
\frac{\dot{w_1}}{{w_1w_2}} - 2\frac{a}{w_2} - 2\frac{B}{w_1} 
-\log(w_1) \frac{{A}}{w_2} \right ]$ \\
& $ \hskip0.3in - 2b \exp(At) \log(w_2) + \frac{d}{dt}G(t, w_1, w_2)$ \\ \hline
\end{tabular}\end{center}

$^*$ $G(t, w_1, w_2)$ is an arbitrary function (see Eq. \ref{s4eq6}).\\

Table 4.  Non-standard Lagrangians for the population dynamics models 
             derived by using the Jacobi last multiplier method (after 
             Nucci \& Tamizhmani [49]).\\

The Lagrangians given in Table 4 are non-standard as they depend explicitly 
on time and neither kinetic nor potential energy terms can be identified.  By 
substituting these NSLs into the E-L equation, the coupled and nonlinear 
first-order equations of motion for the Lotka-Volterra, Verhulst and Gompertz
models of Table 1 and the Host-Parasite model of Table 2 are obtained.  
Therefore, the forms of these NSLs are complicated as they are explicit
functions of both dependent variables, the constant coefficients and time
as well as an arbitrary function $G(t, w_1, w_2)$ that is required to 
correctly apply Noether's theorem [94].  

The NSLs presented in Table 4 are the same as originally found by 
Trubatch and Franco [73], whose method was similar to the Jacobi
last multiplier method and some of their functions play the same 
role as the Jacobi last multiplier as pointed out by Nucci and 
Tamizhmani [49]; the NSL for the Lotka-Volterra model was also
found independently by Saha and Talukdar [25].  The similarities 
between the methods were not obvious to Trubatch and Franco, 
who not being familiar with the method could not get any NSL for 
the Host-Parasite model.  The full derivation of the NSLs given in 
Table 4 was presented by Nucci and Tamizhmani [49], who also 
demonstrated how to apply the method to higher (second and 
fourth) order ODEs.  Thus, the main advantage of the Jacobi last 
multiplier method is that the method gives the NSLs but it does 
not require introducing any forcing functions and, in addition, 
the method gives the same NSLs as those obtained before in
[72,73,75]. 

\subsection{El-Nabulsi method and its non-standard Lagrangians}

An important class of NSLs called the El-Nabulsi Lagrangians contains 
exponential, logarithmic, or different power-laws of standard Lagrangians 
[37] and, to the best of our knowledge, they have not yet been applied 
to any population dynamics models.  The general method to obtain the
NSLs is described in different papers by El-Nabulsi [37-42], where some
applications of these Lagrangians are also presented and discussed.  In 
the following, two such NSLs are considered and their applications to 
the models is proposed.  Since the NSLs require the SLs and since the
SLs are defined here as the Lagrangians that do not depend explicitly 
on time, the results presented below follow [39]; however, they are 
given for the dependent variables $w_i (t)$, $i$ = 1 and 2, for the 
SLs that are not functions of time.  

For the exponential NSL, the action (see Eq. \ref{s2eq1}) becomes 
\begin{equation}
\mathcal{A} (\dot w_i, w_i) = \int \e^{L_{s,i} (\dot w_i, w_i)} dt\ ,
\label{s5eq1}
\end{equation}
and then $\delta \mathcal {A} =0$ gives the following E-L equations
\begin{equation}
{d\over{dt}}\left({{\partial L_{s,i}} \over{\partial{\dot w_i}}}\right) 
- {{\partial L_{s,i}}\over{\partial w_i}} + \left [ \dot w_i \frac{\partial 
L_{s,i}}{\partial w_i} + \ddot w_i \frac{\partial L_{s,i}}{\partial \dot w_i} 
\right ] \left({{\partial L_{s,i}} \over{\partial{\dot w_i}}}\right) = 
{\mathcal F}_{i}(w_i,\dot w_{i}) e^{2I_{\alpha,i}(w_i)}\ .
\label{s5eq2}
\end{equation}
However, for the power-law $(1 + \gamma)$ NSL, the action becomes 
\begin{equation}
\mathcal{A} (\dot w_i, w_i) = \int L_{s,i}^{1 + \gamma} (\dot w_i, w_i) dt\ ,
\label{s5eq3}
\end{equation}
where $\gamma$ is  any real number, and $\delta \mathcal {A} =0$ gives 
\begin{equation}
{d\over{dt}}\left({{\partial L_{s,i}^{1+\gamma}}\over{\partial{\dot w_i}}}
\right) - {{\partial L_{s,i}^{1+\gamma}}\over{\partial w_i}} +  \left [ 
\dot w_i \frac{\partial L_{s,i}^{1+\gamma}}{\partial w_i} + \ddot w_i 
\frac{\partial L_{s,i}^{1+\gamma}}{\partial \dot w_i} \right ] \left({{
\partial L_{s,i}^{1+\gamma}} \over{\partial{\dot w_i}}}\right)= 
{\mathcal F}_{i}(w_i,\dot w_{i}) e^{2I_{\alpha,i}(w_i)}\ .
\label{s5eq4}
\end{equation}
Note that the E-L equations are modified as compared to their original 
forms [39] to account for the dissipative forces in Eq. (\ref{s2eq20})
that are defined below this equation.

To apply these results to the population dynamics models given in 
Tables 1 and 2, it is necessary to use the SLs given in Table 3 and 
calculate the El-Nabulsi Lagrangians $\e^{L_{s,i} (\dot w_i, w_i)}$ 
and $L_{s,i}^{1+\gamma} (\dot w_i, w_i)$, which are required
to find the actions given by Eqs (\ref{s5eq1}) and (\ref{s5eq3}), 
respectively.  However, in order to obtain the equations of motion 
for the models, it is sufficient to substitute the SLs from Table 3 
into the E-L equations given by Eqs (\ref{s5eq2}) and (\ref{s5eq4}), 
respectively; the reason is that the E-L equations already account for 
the specific forms of the El-Nabulsi Lagrangians.  More detailed studies 
are needed to explore implications of these NSLs on the population 
dynamics, specifically, on conserved and non-conserved quantities, 
model's hidden symmetries, and others.

\subsection{Direct method and its non-standard Lagrangians}

The equation of motion given by Eq. (\ref{s3eq1}) and used to obtain 
the SLs for the population dynamics models contains the term $\gamma_{i}
(w_{i})w_{i}$ that limits construction of NSLs to only special dynamical 
systems for which strong relationships between its coefficients in the 
equations of motion exist (e.g., [20-23,  57,58]).   Following [32], 
Eq. (\ref{s3eq1}) can be written as 
\begin{equation}
\ddot w_{i}+\alpha_{i}(w_{i})\dot w_{i}^2 = {\mathfrak F}_{i}(w_i,
\dot w_{i})\ ,
\label{s6eq1}
\end{equation}
where ${\mathfrak F}_{i}(w_i,\dot w_{i}) = {\mathcal F}_{i}(w_i,
\dot w_{i}) - \gamma_{i}(w_{i})w_{i}$.  Despite the existence of 
the nonlinear damping term $\alpha_{i}(w_{i})\dot w_{i}^2$ in 
the equations of motion, the LHS of Eq. (\ref{s6eq1}) is conservative 
[20] and it describes oscillations of the population of species with 
respect to its equilibrium.  The equilibrium and the resulting 
oscillations are modified by the force-like term on the RHS of 
the equation.

The direct method of construction of $L_{ns,i}(\dot w_i,w_i)$
requires substitution of the NSL given by Eq. (\ref{s2eq5}) 
into the E-L equation and evaluating the functions $F(w_i)$,
$G(w_i)$ and $H(w_i)$ as shown in [20].  The result is 
\begin{equation}
L_{ns,i}(\dot w_i,w_i)=\frac{1}{\dot w_i e^{I_{\alpha,i} (w_i)} 
+ C_{o,i}}\ ,
\label{s6eq2}
\end{equation}
where the constants $C_{o,i}$ replace the function $G (x)$ in Eq. 
(\ref{s2eq5}), and these constants can have any real values, which
are not required to be determined; in addition, the integral 
$I_{\alpha,i} (w_i)$ is given by Eq. (\ref{s3eq4}).  Substitution
of $L_{ns,i}(\dot w_i,w_i)$ into the following E-L equation
\begin{equation}
{d\over dt}\biggr({\partial L_{ns,i}\over\partial\dot w_i}\biggr)-
{\partial L_{ns,i}\over\partial w_i} = {\mathfrak F}_{i}(w_i,\dot 
w_{i}) e^{2I_{\alpha,i}(w_i)}\ ,
\label{s6eq3}
\end{equation}
gives the required equation of motion (see Eq. \ref{s6eq1}).
The NSLs obtained from Eq. (\ref{s6eq2}) exist for any integrable 
$\alpha_i (w_i)$ regardless of the forms of $\beta_i (w_i)$ and 
$\gamma_i (w_i)$ [32], and they are presented in Table 5.

\renewcommand{\arraystretch}{1.1}
\begin{center}
\begin{tabular}{@{}lll@{}} 
\hline
{\bf Models}	&{\bf Non-standard Lagrangians}\\ 
\hline
Lotka-Volterra  &	$L_{ns,1} = [\dot{w_1} w_1^{-1} + C_{o,1}]^{-1}$\\
			&	$L_{ns,2} = [\dot{w_2} w_2^{-1} + C_{o,2}]^{-1}$\\ \hline
            
Verhulst  	    &	$L_{ns,1} = [\dot w_1 w_1 ^ {-(1+b)} + C_{o,1}]^{-1}$\\ 
			&	$L_{ns,2} = [\dot w_2 w_2 ^ {-(1+B)} + C_{o,2}]^{-1}$\\ \hline 

Gompertz  	    & 	$L_{ns,1} = [\dot{w_1} w_1^{-1} + C_{o,1}]^{-1}$\\ 
			&	$L_{ns,2} = [\dot{w_2} w_2^{-1} + C_{o,2}]^{-1}$\\ \hline 
   
Host-Parasite	&	$L_{ns,1} = [\dot w_1  w_1^{-1}e^{B \over w1} + C_{o,1}]^{-1}$\\
			&	$L_{ns,2} = [\dot{w_2} w_2^{-2} + C_{o,2}]^{-1}$\\ \hline 
   
SIR  		    &	$L_{ns,1} = [\dot{w_1} w_1^{-1} + C_{o,1}]^{-1}$\\ 
	       	&	$L_{ns,2} = [\dot{w_2} w_2^{-1} + C_{o,2}]^{-1}$\\ \hline
\end{tabular}\end{center}%

Table 5. Non-standard Lagrangians for the population dynamics models 
             derived by using the direct method (after Pham \& Musielak [32]).\\

An interesting result is similarity of the NSLs given in Table 5; all the NSLs
contain the same term $\dot w_i w_i^{-1}$, which is only slightly modified 
for the Verhulst and Host-Parasite models, and constants $C_{o,i}$ that are 
arbitrary and do not have to be evaluated.  The presented NSLs are much 
simpler than those given in Table 4, which were obtained by the Jacobi last 
multiplier method.  This simplicity is caused by the fact that some terms in
the equations of motion come from the forcing functions introduced for the 
population dynamics models.  The main advantage of the NSLs given in 
Table 5 is that they can be easily converted into null Lagrangians, whose 
gauge functions can also be easily found as shown in the following section.  

\section{Null Lagrangians and gauge functions for population dynamics models}

\subsection{Construction of null Lagrangians}

As demonstrated in Section 2.4, null Lagrangians (NLs) are easy to construct 
as the total derivative of any scalar function $\Phi_i (w_i,t)$ gives $L_{n,i} 
(\dot w_i,w_i, t)$ that leads to an equation of motion when substituted 
into $d L_{n,i} / dt = 0$ (see Eq. \ref{s2eq8}), which plays the same role
for the NLs as the E-L equation plays for the SLs and NSLs.  However, the 
equation of motion obtained this way is arbitrary and typically unrelated to 
the equation of interest [57].  Interestingly, there are a very few dynamical 
systems, whose equations of motion can be obtained from NLs that are 
derived specifically for those equations, and such a procedure is described 
in [58].  The procedure was used in [32] to find first NLs for the population 
dynamics models and to demonstrate that such NLs give the correct equations 
of motion for the models as it is now shown.

\subsection{Resulting null Lagrangians and their gauge functions}

The first NLs and their gauge functions for the population dynamics models
are presented in Table 6 following [32].  Comparison of the results given 
in Table 6 to those from Table 5 shows that the NLs are inverse of the 
NSLs as originally demonstrated in [57].  It is important to point out that 
the NLs presented in Table 6 give the equations of motion for the models
when they are substituted into 
\begin{equation}\
\frac{d L_{n,i}}{dt} = {\mathfrak F}_{i}(w_i,\dot w_{i}) e^{I_{\alpha,i}
(w_i)}\ ,
\label{s7eq1}
\end{equation}
where $I_{\alpha ,i}$ is given by Eq. (\ref{s3eq4}); note the modification
of the original condition $d L_{n,i} / dt = 0$ presented in [56].  Having 
obtained the NLs, then the corresponding GFs are determined by 
\begin{equation}\
L_{n,i}(\dot w_i, w_i) = \frac{\partial \Phi_{i}(\dot x, x,t)}{\partial t} 
+ \dot w_i  \frac{\partial \Phi_{i}(\dot w_i, w_i)}{\partial w_i}\ ,
\label{s7eq2}
\end{equation}
and they are given in Table 6.

\renewcommand{\arraystretch}{1.1}
\begin{center}
\begin{tabular}{@{}lll@{}} 
\hline
{\bf Models}	&{\bf Null Lagrangians } &{\bf Gauge functions}\\ 
\hline
Lotka-Volterra  &	$L_{n,1} = \dot{w_1} w_1^{-1} + C_{0,1}$ 
                &   $\Phi_{1} = \ln \vert w_1 \vert + C_{0,1} t$ \\
			&	$L_{n,2} = \dot{w_2} w_2^{-1} + C_{0,2}$ 
                &   $\Phi_{2} = \ln \vert w_2 \vert + C_{0,2} t$\\ \hline
            
Verhulst  	    &	$L_{n,1} = \dot w_1 w_1 ^ {-(1+b)} + C_{0,1}$ 
                &   $\Phi_{1} = - b^{-1} w_1^{-b} + C_{0,1} t$ \\
			&	$L_{n,2} = \dot w_2 w_2 ^ {-(1+B)} + C_{0,2}$ 
                &   $\Phi_{2} = - B^{-1} w_1^{-B} + C_{0,2} t$\\ \hline

Gompertz  	    & 	$L_{n,1} = \dot{w_1} w_1^{-1} + C_{0,1}$ 
                &   $\Phi_{1} = \ln \vert w_1 \vert + C_{0,1} t$ \\
			&	$L_{n,2} = \dot{w_2} w_2^{-1} + C_{0,2}$ 
                &   $\Phi_{2} = \ln \vert w_2 \vert + C_{0,2} t$\\ \hline
   
Host-Parasite$^*$&	$L_{n,1} = \dot w_1  w_1^{-1}\e^{B \over w1} + C_{0,1}$ 
                &   $\Phi_{1} = - Ei ({B \over w_1}) + C_{0,1} t$ \\
			&	$L_{n,2} = \dot{w_2} w_2^{-2} + C_{0,2}$ 
                &   $\Phi_{2} = - w_2^{-1} + C_{0,2} t$\\ \hline
                
SIR  		    &	$L_{n,1} = \dot{w_1} w_1^{-1} + C_{0,1}$ 
                &   $\Phi_{1} = \ln \vert w_1 \vert + C_{0,1} t$ \\
	       	&	$L_{n,2} = \dot{w_2} w_2^{-1} + C_{0,2}$ 
                &   $\Phi_{2} = \ln \vert w_2 \vert + C_{0,2} t$ \\ \hline
\end{tabular}\end{center}%

$^*$ $ Ei ({B \over w_1}) = \int_{- \infty}^{B/w_1} (\e^{\eta}/ {\eta}) 
d \eta$ 
is the exponential integral.\\

Table 6. Null Lagrangians and their gauge functions for the population 
            dynamics models (after Pham \& Musielak [32]).\\

The presented NLs in Table 6 and their GFs are first null Lagrangians
and gauge functions obtained for biological systems [32]; their roles
and implications for the systems, specifically, for the population dynamics
models and their equations of motion, still require further investigation. 
The results of Table 6 show that the GFs for the symmetric models are 
given by logarithmic functions of the dependent variables and that they
are linear functions of time.  By comparing the GFs to those obtained for
different oscillatory dynamical systems [96], it is seen that the derived 
GFs form a new class of gauge functions.  

Similarly, the GFs obtained for the asymmetric models are also 
very different from those for the symmetric models.  The GFs for 
the Verhulst models are power-laws of their dependent variables.  
However, the most unusual form has the GF for the Host-Parasite 
model for its variable $w_1$, the GF is expressed as the exponential 
integral $ Ei ({B \over w_1})$,which is a novel gauge function; similar 
term is also present in the SL found for this model (see Table 3).  The 
above results show that the similarities and differences can be used to 
make distinctions between the symmetric and asymmetric models as 
well as to classify these models. 

\section{Applications of obtained Lagrangians to population dynamics}

\subsection{Equations of motion and their symmetries}

The derived SLs, NSLs and NLs for the population dynamics models
are very different, and yet they give the same equations of motion 
when substituted into the E-L equations, whose forms may vary 
from one method to another.   In general, the obtained SLs may 
be considered to be the most basic Lagrangians for the model, 
with the NSLs and NLs being alternative Lagrangians.  Nevertheless, 
all the derived Lagrangians are important without exceptions because 
different Lagrangians may show different symmetries.  In other words, 
the NSLs and NLS may display different symmetries than those shown 
by the SLs.  The main reason is Noether's theorem [95], and the fact 
that NSLs may reveal a symmetry that is hidden and not display by 
the SLs (e.g., [46]). 

Let us explain the last point by starting with the following statement: 
'Noether's theorem applied to different Lagrangians yields different 
Noether's symmetries and thus different conservation laws. Therefore, 
as many Lagrangians as possible for an equation must be found and 
then Noether's theorem can identify those physical Lagrangians that 
either yield the missing conservation laws or successfully lead to 
quantization', written by Nucci and Leach [47].  

As the above citation shows, the obtained SLs, NSLs and NLs can be 
used to determine Noether [97-99], non-Noether [100-102], and other 
[17-19,103] symmetries of Lagrangians and the equations of motion.  
The fact that Noether symmetries of SLs remain the same whether 
any NL is added to them or not is well-known (e.g., [97]).  However, 
a SL and SL+NL yield their two corresponding non-Noether symmetries 
in a unique way [98].  In general, SLs, NSLs and NLs possess less 
symmetry than the equations of motion resulting from them due 
to assumptions on which the Noether theorem [17-19] is based.  

Now, in previous studies of the Lagrangian formalism in physics 
and biology, the gauge functions that generate null Lagrangians 
were typically omitted because the NLs do not generate any 
equation of motion when substituted into the E-L equation. 
Extensive studies of the GFs and NLs in mathematics (e.g., 
[59-65]) were not directly related to the Lagrangian formalism;
however, some studies in physics have considered possible 
applications of the GFs and NLs to specific physical problems (e.g., 
[66-77,96]).  A recent work [57,58] demonstrated roles of 
the GFs and NLs in obtaining equations of motion was followed 
in [32], where first NLs for the population dynamics models were
obtained, and it was suggested that these gauge functions may 
play important roles in relating the SLs, NSLs and NLs for the 
models.  It was also pointed out that more studies of the GFs 
are needed to fully determine their roles in biological systems.

\subsection{Identifying conserved and nonconserved quantities}

In studies of dynamical systems in physics, one of the most important 
tasks is to identify quantities that remain constant in time.  This can
be done for equations of motion or for Lagrangians that give these
equations.  However, it is required that such Lagrangians are not 
explicit functions of time.  Among the Lagrangians presented in this 
review paper, only the NSLs obtained by the Jacobi last multiplier
method depend on time.  The derived SLs and other NSLs and NLs 
are not explicit In functions of time because their time dependence 
is hidden in the forcing terms present in the E-L equations.  In the 
following, we describe the results presented in [23] and make some 
general comments about their extension to the NSLs and NLs.    

The fact that the derived SLs do not depend explicitly on time (see 
Table 3), their total energy $E_{tot} = E_{kin} + E_{pot}$ must 
be conserved, which can be shown by calculating the energy 
function $E_{fun, i}$ [6,7] for the SLs given in Table 3.  The
result is 
\begin{equation}
E_{fun, i} (\dot w_i,w_i) = \dot {w}_i \frac{\partial L{s,_i}}{\partial 
\dot {w_i}} - L_{s,i} (\dot w_i , w_i)\ ,
\label{s8eq1}
\end{equation}
which gives $E_{fun, i} (\dot w_i,w_i) = E_{tot, i} (\dot w_i,w_i) = 
E_{kin, i} (\dot w_i,w_i) + E_{pot, i} (w_i)$ = constant for all 
the considered models.  Then, the energy function can be used 
to derive equations of motion [6,7] by using 
\begin{equation}
\frac{dE_{fun, i}}{dt} = - \frac{\partial L_{s,i}}{\partial t}\ ,
\label{s8eq2}
\end{equation}
which also gives $E_{fun, i} (\dot w_i,w_i) = E_{tot, i} (\dot w_i,w_i)$ 
= constant because the derived SLs do not depend explicitly on time. 

The origin of the conserved quantity $E_{tot, i}$ can be explained 
by the fact that the SLs describe only the homogeneous parts of the 
ODEs that represent the models, namely, substitution of the SLs into
the E-L equation gives  
\begin{equation}
\ddot w_{i}+\alpha_{i}(w_{i})\dot w_{i}^2+\gamma_{i}(w_{i})
w_{i}=0\ ,
\label{s8eq3}
\end{equation}
which describes the populations of interacting species that oscillate in time.
The oscillatory behavior may be surprising because of the presence of the  
and despite the presence of the quadratic damping $(w_i^2)$ term the
equation of motion; however, previous studies showed that this is indeed
the case [20,21].

The total energy determined above remains only conserved for the system
given by Eq. (\ref{s8eq3}), which does not account for the force-like function 
${\cal F}_i (\dot w_i, w_i)$ that includes the linear damping term (see 
Eqs \ref{s2eq7} and \ref{s3eq1}).  The definition of ${\cal F}_i (\dot 
w_i, w_i)$ seems to be natural because the damping term is equivalent to 
a null Lagrangian, which means that no SL can account properly for such a 
NL [23].  Thus, the effects of ${\cal F}_i (\dot w_i, w_i)$ on the 
conserved total energy must be now discussed.

The main effect of ${\cal F}_i (\dot w_i, w_i)$ on the system's oscillatory 
behavior is that the system reaches the equilibrium either faster, due to damping,
or diverges from it, due to energy being added.  Since ${\cal F}_i (\dot w_i, 
w_i)$ for all the models, except the Gompertz model, is linear in $\dot w_i$, the 
forcing function can be expressed as ${\cal F}_i (\dot w_i, w_i) = f_i (w_i) 
\dot w_i$, where $f_i (w_i)$ accounts for all the terms that depend exclusively on 
$w_i$.  This allows defining the Rayleigh dissipative function [6] in the following 
from
\begin{equation}
R_i (\dot w_i, w_i) = \frac{1}{2} f_{e, i} (w_i) \dot w_i^2\ , 
\label{s8eq4}
\end{equation}
where $f_{e, i} (w_i) = f_i (w_i) e^{2I_{\alpha}(w_i)}$, and write
the E--L equation in the following form
\begin{equation}
{d\over{dt}}\left({{\partial L_i }\over{\partial {\dot w_i}}}\right) - 
{{\partial L_i}\over{\partial w_i}} = \frac{\partial R_i}{\partial \dot w_i}\ .
\label{s8eq5}
\end{equation}
Note that the sign of the term on the RHS of the above equation is determined 
by the sign of the function ${\cal F}_i (\dot w_i, w_i)$ or $f_i (w_i)$, 
and it is different for different population dynamics models.  In general, the 
minus sign means a 'damping force' and the plus sign means a 'driving force'.  
For the Gompertz model, Rayleigh's force cannot be defined because $\dot w_i$ 
is nonlinear and logarithmic, which makes it impossible to separate from the 
remaining terms of the expression.  

For those population dynamics models that the Rayleigh's function can be 
defined, the energy function can be calculated by using  
\begin{equation}
\frac{dE_{fun, i}}{dt} = - \frac{\partial L{s_i}}{\partial t} + R_i (\dot w_i, 
w_i)\ ,
\label{s8eq6}
\end{equation}
or by
\begin{equation}
\frac{dE_{fun, i}}{dt} = R_i (\dot w_i, w_i)\ ,
\label{s8eq7}
\end{equation}
because $L_{s,i} (\dot w_i, w_i) \ne L_{s,i} (t)$.   This is an important 
result because integration of Eq. (\ref{s8eq7}) in time gives changes of
$E_{fun, i} (\dot w_i,w_i)$ and shows their effects of damping or driving 
on otherwise oscillatory behavior of the population dynamics models; 
however, it must be noted that in order to perform the integration, the 
solutions for $w_i (t)$ must be known. 

Since NSLs and NLs presented in Tables 5 and 6, respectively, are 
also not explicit functions of time, the above procedure can be used 
to determine conserved quantities for the population dynamics models.
The conserved quantities will not be identified as total energy because 
the terms in those NSLs and NLs do not represent energy-like terms.
Nevertheless, the calculated conserved quantities will determine the 
equilibria for the population dynamics models and the forcing function
${\mathfrak F}_i (\dot w_i, w_i)$ will be used to find deviations from 
the equilibria, similar as it was done above for the SLs.  This work still
remains to be done and when accomplished it may give new conserved
quantities that may allow defining new equilibria for the models.

\subsection{Oscillatory behavior of the populations}

The fact that the oscillatory behavior in the population dynamics
models can be separated from the driving and damping functions
is the main result presented in [23,35], in addition, to the SLs,
NSLs and NLs and their GFs that are derived and discussed there. 
The result is important as it allows for independent studies of the 
oscillations, and then using Eq. (\ref{s8eq7}) to determine the 
effects of damping or driving on these model oscillations.  In 
other words, this separation allows finding the period of oscillations 
by considering the minimization of the following quantity [73,104]
\begin{equation}
Z_i [w_i (t)] = \int^P_0 [ L_i (\dot w_i, w_i) + E_{tot, i} (\dot w_i, 
w_i)] dt\ ,
\label{s8eq8}
\end{equation}
where $P$ is the period of the oscillations, which is an unknown, 
and it is required that $E_{tot, i} (\dot w_i, w_i)]$ is conserved
If the solutions for $w_i (t)$ are known and the integration can 
be performed either analytically or numerically, then the value  
of $P$ can be determined and verified by experimental data 
collected for the population dynamics models.

There are three main population dynamics insights emerging 
from the presented results [23]: (i) the oscillatory and damping 
or driving functions of the models can be treated separately and 
studied independently; (ii) the theoretically predicted period of
of oscillations can be verified by comparing it to experimental data; 
and (iii) there is a procedure that allows accounting for the damping 
or driving effects on the already established oscillatory behavior in 
the population dynamics models.    

All the developed procedures and the presented results were 
obtained for the five selected population dynamics models, but
they can be easily extended to other population dynamics models
as well as to a broad range of different ecological and biological 
systems.

\subsection{Other real-world applications}

The population dynamics models considered in this paper have also 
been applied to real-world ecological scenarios, and some 
recent work has involved the Lotka-Volterra model to study the 
marine phage population dynamics (e.g., [105]), to investigate
changes in density of a population in community ecology (e.g., 
[106]), and to also explore stability of this model with time-varying 
delays (e.g., [107]).  There are also different applications of other 
models (e.g., [108]), specifically, with the recent applications of 
the SIR model to the COVID pandemic (e.g., [109,110]). 

New insights into those studies can be made by using the constructed
SLs, NSLs and NLs by the method described in this paper.  The reason 
is that the SLs would allow identifying the total energy-like terms and 
establish their conservation in the models.  Moreover, the NSLs and 
NLs could be used to determine other conserved quantities.  Then, 
the equilibria for the models could be found and their periods of 
oscillations.  The analysis of the forcing functions would give 
information about their effects on the oscillatory behavior of 
the populations described by the models.   

Finally, let us point out that the presented methods to derive the 
SLs, NSLs, NLs and GFs, and the procedure developed to determine 
the equilibria, periods or oscillations and the effects of driving and 
dissipative forces is powerful and can be used to any ecological or 
biological system, which is represented mathematically by either 
first-order or second-order ordinary differential equation that is 
nonlinear, has both linear and quadratic damping terms, and a 
given driving force; the methods and procedure also work in 
special cases when one or more terms are not present.  

\section{Perspectives of Lagrangian formalism in biology}

The existence of Lagrangian for a given equation of motion is one of
the requirements in theoretical physics for the equation to be called 
fundamental (e.g., [111]).   Thus, all the fundamental equations of 
modern physics are derived from the Lagrangian formalism (e.g.,  
[8-11]).  In the view of the authors of this review paper, the main 
goal of future work on Lagrangian formalism in biology should be 
making the formalism to be the basis for theoretical biology.   To 
achieve this goal, significant future work is needed, specifically,
by extending the methods to find Lagrangians described in this 
paper to more complex biological systems.   Since the presented
methods are valid for one-dimensional (changes in time only) and 
two-species population dynamics models, and since many reali
tic 
biological systems are multi-dimensional (changes in time and space) 
and multi-species, it is required that either currently existing methods 
are extended to such systems, or that new methods are developed 
to account properly for the system's complexity and its interactions 
with the environment.   

As shown in this paper, knowledge is Lagrangians for different 
biological systems is essential to determine physical properties 
of the systems without formally solving their equations of motion.
Such properties include, but are not limited to, the system's 
equilibrium, system's oscillations with respect to its equilibrium,
and the effects of dissipative and other forces on the systems.  
The previous work in this area can only be recognized as 
preliminary, thus, more future studies are needed to develop 
procedures that are applicable to complex biological systems.  
After such procedures are developed, they could also be used 
to investigate dynamical stability of the systems, including 
stability of their equilibrium points.  The stability analysis of 
biological systems could be used to derive optimal control 
strategies for managing populations; with an appropriate 
cost function, the analysis could determine the best strategies 
to achieve desired population levels.  Another application of 
the stability analysis could involve studies of small deviations 
(perturbations) from their equilibria, which are caused by 
some small effects of the environment on the systems. 
  
It was briefly pointed out in Section 8.1 that different equations
of motion may have different symmetries, and that symmetries 
of such equations may be different than symmetries of their 
Lagrangians (e.g., [17-19]).  In general, symmetries can be 
classified as Noether's [97-99], non-Noether's [100-102], and 
others [17-19,103].   There are no studies of any of these 
symmetries for the population dynamics models; the only 
exception is the paper by Nucci and Sanchini [50], who 
investigated symmetries and conservation laws of an Easter 
Island population model.  Similar future work is required for 
other population dynamics models and for other biological 
systems.  This future work must involve both the equations 
of motion as well as their standard, non-standard and null
Lagrangians.  More detailed future studies of gauge functions 
are also necessary as these functions may give new physical 
insights into the behavior and evolution of biological systems
(e.g., [32,58]).

Symmetries of equations of motions and their Lagrangians are 
directly related to Lie groups (e.g., [11]).  Let $L [\dot w_i (t), 
w_i (t), t]$ be a Lagrangian, $w_i$ be one of the dependent 
variables in the population dynamics models, $M$ be a configuration 
manifold, and $TM$ be a tangent bundle associated with this manifold;
then $L: TM \rightarrow \mathcal{R}$, and $L$ is defined on $TM$ [7].  
Now, let $G$ be a manifold associated with a given Lie group $G$, so 
that $L: TG \rightarrow \mathcal{R}$.  Then, the Lagrangian $L [\dot 
w_i (t), w_i (t), t]$ is invariant, and its invariance is strongly related to
its Lie group [11].  In general, for every known Lagrangian its invariance 
with respect to rotations and translations may indicate the presence of 
the underlying Lie group [112,113]; in other words, the group may be 
identified by investigating the Lagrangian invariance.  It is also known 
(e.g., [97-99]) that the invariance of $L [\dot w_i (t), w_i (t), t]$ 
guarantees that the original equation of motion preserves the same 
invariance as the Lagrangian from which the equation is obtained.  
Future studies of symmetries and their underlying Lie groups in 
biological systems would lead to a better understanding of the 
interactions between the species involved, as well as their 
evolution in time and space that includes interactions with 
the environment. 

\section{Conclusions}

In physics, a classical particle or example of such particle that evolves
in time and is described by an ordinary differential equation is called a
dynamical system.  The concept has been extended to biology, where the 
'particles' are replaced by living organisms that evolve over time due 
to their mutual interactions and interactions with the environment.
There are different methods to establish an equation of motion for a 
given dynamical system, and one of the most powerful method is to 
know a function called Lagrangian.  If such a function is known, then
the equation of motion can be derived from the Principle of 
Stationary Action that forms the basis of the calculus of variations.
The procedure of finding Lagrangians for given equations of motion
is called the inverse problem of the calculus of variations or the 
Lagrangian formalism.

In this paper, a broad review of the Lagrangian formalism in biology
is presented in the context of its historical and modern developments.  
The methods to obtain standard, non-standard and null Lagrangians, 
and the gauge functions corresponding to the latter, are reviewed 
in details and applied to the following population dynamics models: 
the Lotka-Volterra, Verhulst, Gompertz, Host-Parasite and SIR models.  
It is shown that the Lagrangians formalism allows finding different 
Lagrangians for the same model, and that each one of these 
Lagrangians gives the same equation of motion, which means 
that the Lagrangians derived for a given model are equivalent.  

The presented results also demonstrate that the variety of Lagrangians 
derived for the same population dynamics model becomes an important 
tool to investigate conservation laws and symmetries for the models by 
using Noether's theorem and Lie groups.  The reason is that different 
Lagrangians may display different conservation laws and symmetries,
or in other words, different Lagrangians may allow finding hidden 
symmetries and laws and, therefore, lead to better understanding 
the time evolution of such dynamical systems.   The paper reviews
methods to determine conserved and nonconserved quantities for 
the models, their equilibria and period of oscillations around these 
equilibria.  An important result discussed in the paper is the fact 
that dissipative forces can be separated from the system's oscillatory
behavior, which leads to new insights into the studies of the models.

We hope that this paper presents a balanced view of the Lagrangian 
formalism in biology, and it clearly demonstrates that the formalism 
is already established, and that it can be used to derive equations of 
motion for the population dynamics models as well as for many other 
biological systems.   We also hope that our specific selection of topics
covered in the paper well reflects the previous and current research 
in the field, and they can serve as a guide to the major achievements
in the field.  The main purpose of this review paper will be accomplished
if it serves as a guide to the major achievements in the field, and as 
an inspiration to scientists and students for opening new frontiers of 
research that would make the Lagrangian formalism to become a 
foundation of theoretical biology as it is the foundation of theoretical 
physics.  

Finally, let us point out that our reference list is only a small subset 
of all papers and books published by mathematicians and physicists 
on the Lagrangians formalism.  Our preference has been to cite all
references directly relevant to the population dynamics and biology,
and also those written by physicists and mathematicians that are 
directly relevant to the covered topics; however, we do respect that 
other scientists working in the field may have different opinions in 
this matter.

\bigskip
\noindent
{\bf Acknowledgments:} We would like to thank R. Das, L.C. Vestal and 
A.L. Segovia for discussing with us different aspects of the Lagrangian 
formalism.\\

\end{document}